\documentclass[a4paper,fleqn,usenatbib]{mnras}
\usepackage{hyperref}
\usepackage[T1]{fontenc}
\usepackage{epsfig}
\usepackage{ae,aecompl}
\usepackage{graphicx}    % Including figure files
\usepackage{amsmath}     % Advanced maths commands
\usepackage{amssymb}     % Extra maths symbols
\usepackage{gensymb}     % Extra symbols 
\usepackage{pdflscape}
\usepackage{comment}
\usepackage{subcaption}
\usepackage{breakurl}
\usepackage{arydshln}
\usepackage{newtxtext,newtxmath}

\hypersetup{pdftitle=Spectro-Temporal Study of TXS 1700+685}

\title[Spectro-Temporal Study of TXS 1700+685]{BROADBAND SPECTRO-TEMPORAL STUDY ON BLAZAR TXS 1700+685}
\author[Banerjee, Nandi, Prince, Khatoon \& Bose]{
Anuvab Banerjee$^{1}$\thanks{E-mail: anuvab.banerjee@bose.res.in},
Prantik Nandi$^{1,5}$\thanks{E-mail: prantiknandiphy@gmail.com, prantik@prl.res.in},
Raj Prince$^{2}$\thanks{E-mail: rajprince59.bhu@gmail.com},
Rukaiya Khatoon$^{3,4}$\thanks{E-mail:  rukaiyakhatoon12@gmail.com},
Debanjan Bose$^{1}$\thanks{E-mail: debanjan.tifr@gmail.com}
\\
$^{1}$S. N. Bose National Centre for Basic Sciences, Block JD, Salt Lake, Kolkata 700106, India\\
$^{2}$Center for Theoretical Physics, Polish Academy of Sciences, Lotnik\'{o}w 32/46, Warsaw, Poland\\
$^{3}$Inter-University Center for Astronomy and Astrophysics, Post Bag 4, Ganeshkhind, Pune-411007, India\\
$^{4}$Tezpur University, Napaam-784028, Assam, India \\
$^{5}$Physical Research Laboratory, Navrangpura, Ahmedabad, 380009, India}

\date{Accepted XXX. Received YYY; in original form ZZZ}

\pubyear{2022}

\begin{document}
\label{firstpage}
\pagerange{\pageref{firstpage}--\pageref{lastpage}}
\maketitle

\begin{abstract}
We attempt to present a multiwavelength variability and correlation study as well as detailed multi-waveband spectral characteristics of the May 2021 $\gamma$-ray flare of the blazar source TXS 1700+685.
The multi-wavelength observation from \textit{Fermi}-LAT, \textit{Swift}-XRT/UVOT as well as radio archival data are used for our spectro-temporal investigation. We estimate the variability time-scale of the source from the flux doubling time in different flaring region detected in \textit{Fermi}-LAT observation and the shortest variability time is used to put a constraint on the minimum Doppler factor and on the size of the emission region. We have detected a statistically significant quasi-periodic oscillation feature (QPO) at $\sim$ 17 days. The broad-band emission is satisfactorily represented during its flaring state with a leptonic synchrotron and inverse Compton component.
From the broadband spectral modelling, we observe the external Comptonization of the seed photons originating in the broad line region to be dominant compared to the dusty torus. This is further supported by the fact that the emission region is also found to be residing within the BLR. The equipartition value implies the energy density of the magnetic field in the jet comoving frame is weak, and that is also reflected in the magnetic field and low power corresponding to the magnetic field component of the jet.
In order to produce the high energy hump, we need the injection of a large population of high energy electrons and/or the presence of strong magnetic field; and we observe the later component to be sub-dominant in our case. The flat rising and steep falling profile in the $\gamma$-ray SED as well as the break or spectral curvature at $\sim$ 1 GeV are in commensuration with the flat-spectrum radio quasar (FSRQ) nature of the source.

\end{abstract}


%%%%%%%%%%%%%%%%%%%%%%%%%%%%%%%%%%%%%%%%%%%%%%%%%%

%%%%%%%%%%%%%%%%% BODY OF PAPER %%%%%%%%%%%%%%%%%%

\section{Introduction}
\label{sec:intro}
Blazar sources are a subclass of active galactic nuclei (AGNs) that emit copiously in all frequencies
extending from the radio to very high-energy (more than a few tens of GeV) $\gamma$ ray band, and their relativistic jets are oriented almost along the line of sight (angles less than $\sim 14^{\circ}$) to the observer \citep{urry1995unified}. Owing to the complex characteristics of AGNs as well as the lack of high-resolution instruments, the exact origin of such high energy emission as well as the complete physical mechanism driving such energetic outflowing matter has remained unresolved for several decades. The association between very high energy emission and jet activity became evident after the Energetic Gamma-ray Experiment Telescope (EGRET) observations onboard the Compton Gamma-Ray Observatory (CGRO; \citet{hartman1999}). This has been more comprehensively established by Fermi/LAT (Fermi Gamma-Ray Space Telescope/Large Area Telescope) extensive all-sky survey.  \par 

Another defining characteristic of blazar sources is their large amplitude flux variability across different wavebands which has been used previously to approximate the location and size of the emission regions (see \citet{hovatta2019relativistic} for a review). In many of the cases, the optical flares are found to be correlated with $\gamma$-ray flares \citep{chatterjee2012similarity,carnerero2015multiwavelength}. However, this may not always be the case and $\gamma$-ray flares may not even be associated with an optical counterpart \citep{vercellone2011brightest, macdonald2015through, cohen2014temporal, rajput2019temporal}. Further, the short time scale of high energy $\gamma$-ray flux variation is indicative of the fact that such emission is originated from very compact regions in the jet. \par 

A double hump broad-band spectral energy distribution (SED) is apparent in blazar, where the low energy hump peaks at IR - X-ray
band and the higher-energy component peaks in the MeV-GeV band \citep{1998MNRAS.299..433F,mao2016comprehensive}. The low energy hump is now believed to be originated from the synchrotron emission of the relativistic electrons in the jet. However, the origin of higher energy hump is still debated, and there could be a range of physical mechanisms that contribute to the genesis of this hump. In the leptonic scenario, the high-energy hump is attributed to the inverse Compton (IC) scattering of low-energy photons within the jet. The seed photons for IC scattering can be supplied by the jet itself (synchrotron self-Compton or SSC; \citet{konigl1981relativistic,marscher1985models,ghisellini2009canonical}), or the origin of the seed photons can be external to the jet (external Compton or EC). In EC process,  the seed photons can be supplied by the disc \citep{dermer1993model,bottcher1996gamma,bottcher1997spectral}, the dusty torus \citep{ghisellini2008blazar}, or the broad-line emission (BLR) region \citep{sikora1994comptonization}. \par 

Fermi Gamma-ray Space Telescope onboard the Large Area Telescope (LAT) detected enhanced $\gamma$-ray activity on March 22, 2009 from the blazar TXS 1700+685 \citep{2009ATel.1986....1G}. This source is also known as 4FGL J1700.0+6830 \citep{2020ApJS..247...33A} and is located at redshift z = 0.301 \citep{1997MNRAS.290..380H} with R.A. = 255.038 deg. and Decl. = +68.5019 \citep{petrov2008sixth}. Recently this source has shown renewed $\gamma$-ray activity on March 15, 2021 when the daily averaged gamma-ray flux ($E > 100$ MeV) rising to $(1.1\pm 0.2)\times 10^{-6} \text{ erg cm}^{-2}\text{s}^{-1}$, with the detection of a 29 GeV photon \citep{2021ATel14463....1C}. This daily average flux is a factor $\sim$ 20 more compared to the 10-year average flux reported in the fourth Fermi LAT source catalog Data Release 2 (4FGL-DR2; \citet{ballet2020fermi}). Subsequently on May 13, more prominent flaring activity in $\gamma$-ray was registered by \textit{Fermi}-LAT, when daily average $\gamma$-ray flux rose to $(1.8\pm 0.2)\times 10^{-6} \text{ erg cm}^{-2}\text{s}^{-1}$; a flux increase by a factor $\sim$ 30 relative to the daily average flux. On May 14 and 15, it showed a daily averaged gamma-ray flux ($E > 100$ MeV) $(1.0\pm 0.2)\times 10^{-6} \text{ erg cm}^{-2}\text{s}^{-1}$ \citep{2021ATel14633....1C}. \par 

In this paper, we have attempted to investigate the possible correlation between the different wavebands, as well as the accretion and jet behavior from the multi-wavelength spectral analysis during May 2021 flare, corresponding to which simultaneous \textit{Swift} XRT/UVOT observations are present. The organization of the paper is as follows: In Section 2, we describe the data reduction and analysis techniques; in Section 3, we describe our result on the multi-waveband spectro-temporal analysis, and in Section 4, we draw our conclusions.

%%%%%%------------------------------------------------------------------------------------
     
\section{ Multi-waveband observations and data analysis}
\subsection*{\textit{Fermi}-LAT observations}

Fermi-LAT is a pair conversion $\gamma$-ray detector in orbit with a field of view (FOV) of about 2.4 sr, which is sensitive to photon energies between 20 MeV - 500 GeV \citep{atwood2009large}. The FOV covers $\sim$ 20\% of the sky at any given instant, and it scans the whole sky in every three hours. At 100 MeV, LAT's single-photon resolution is $< 3.5^{\circ}$, but at higher energies ($> 1$ GeV), this resolution improves to $< 0.6^{\circ}$. In order to identify the potential high energy $\gamma$-ray sources within a given region of interest (ROI), a likelihood analysis is needed to be undertaken using an input model using the \texttt{gtlike} task \citep{1979ApJ...228..939C,mattox1996likelihood}. In order to perform our analysis, we have primarily used \texttt{fermitools}\footnote{\url{https://fermi.gsfc.nasa.gov/ssc/data/analysis/documentation/}} package, maintained by the Fermi-LAT collaboration. Usage of \texttt{fermipy}\footnote{\url{https://fermipy.readthedocs.io/en/latest/quickstart.html}} was also required for some specific aspects during spectral analysis \citep{wood2017fermipy}. \par 

We collected 0.1–300 GeV \textit{Fermi}-LAT data within the time-span between MJD 58970 (May 1, 2020) to 59487 (September 30, 2021) for the purpose of our analysis. A circular region of $10^{\circ}$ around the source position was chosen to be the ROI, within which the presence of high energy $\gamma$ ray was searched. As recommended by the \textit{Fermi}-LAT team in \texttt{fermitools} documentation, the user-specified cut \texttt{`evclass=128' and `evtype=3'} were applied for the purpose of the selection of rows from the input event data and the filter \texttt{(DATA\_QUAL>0)\&\&(LAT\_CONFIG==1)} was applied to update the good time interval (GTI) information based on spacecraft specifications. In order to filter any contamination due to the Earth's limp, a zenith angle cut of $90^{\circ}$ was applied. The latest instrument response function \texttt{P8R3\_SOURCE\_V6} was selected for the analysis, and in order to take care of the isotropic and the diffuse background, we used \texttt{iso\_P8R3\_SOURCE\_V3\_v1.txt} and \texttt{gll\_psc\_v21.xml} respectively, available from Fermi
Science Support Center (FSSC). The spectral profiles, as well as parameters, were initialized as per the values published in the 4FGL catalog, and beyond 3 degrees of ROI center, all sources were kept frozen. The $\gamma$-ray spectral profile was compared with three different empirical models:
\begin{enumerate}
	\item Power-law profile, defined as 
	\begin{equation}
		dN(E)/dE = N_0(E/E_0)^{-\alpha},
	\end{equation}
	where $E_0$ is is the energy scaling factor and $\alpha$ is the spectral index.
	\item A log-parabola (LP), defined as
	\begin{equation}
		dN(E)/dE = N_0(E/E_0)^{-\alpha-\beta\ln{(E/E_0)}},
	\end{equation}
	where $E_0$ and $\alpha$ stand as before, and $\beta$ is the curvature index.
	\item A broken power-law (BPL), defined as 
	\begin{equation}
		dN(E)/dE = N_0(E/E_\text{break})^{-\alpha_i},
	\end{equation}
	with $i = 1$ for $E < E_{\text{break}}$ and $i=2$ for $E > E_{\text{break}}$.
\end{enumerate}

\subsection*{X-ray observations of \textit{Swift}-XRT/UVOT}

TXS 1700+685 was observed seven times between 16th May 2021 and 7th June 2021 by \textit{Swift}-XRT/UVOT  (details of the observations are provided in Table 1) with nonzero exposure. We used FTOOLS task \texttt{xrtpipeline} version 0.13.5 to produce cleaned events. The HEASoft version 6.28 and XSPEC version 12.11.1 have been used for the purpose of our analysis. Cleaned event files corresponding to the Photon Counting mode were considered for analysis and circular regions of
radius 20 arcsec centered at the source and slightly away from
the source were chosen for the source and the background
regions respectively. The source and background spectra were produced using \texttt{xselect} task. The redistribution matrix file (RMF) was obtained from the latest HEASARC calibration database (CALDB) and the ancillary response files (ARF) were produced using the task \texttt{ xrtmkarf}. Using \texttt{grppha}, the source event file, background event file, response file, and ancillary response files were tied together and binned in such a way that each bin contains a minimum 10 counts. The spectra were fitted using simple power-law model, with the galactic absorption column density $n_H$ kept fixed at $4.0 \times 10^{20} ~ \text{cm}^{-2}$. \par 

The UV-optical data were obtained via The \textit{Swift}
Ultraviolet/Optical Telescope (Swift-UVOT, \citet{roming2005swift}) which gathers data in six filters: U, V, B, W1, M2, and W2. The source image was extracted from a region of 5 arcsec radius centered at the source and the background image was extracted from a region of 10 arcsec centered away from the source. 

\subsection*{Radio observations}
We did not find any publicly available contemporaneous radio observation corresponding to the 2021 flare. However, for reference, we add non-simultaneous radio data released by the Planck mission as a part of their second release of the Planck Catalogue of Compact Sources (PCCS). Even though the primary objective of Planck mission is to measure the cosmic microwave background (CMB) anisotropy, as a by-product of their operation detailed catalogue of compact sources as well as galactic diffusion maps are released, and we use the data between 2009 August 12 and 2010 November 27 for the purpose of our analysis \citep{2014A&A...571A..28P,2016A&A...594A..26P}. The data is processed by the low-frequency instrument (LFI; 30–70 GHz) and high-frequency instrument (HFI; 100–857 GHz) Data Processing Centres (DPCs) and is downloaded via the publicly available SED Builder tool\footnote{\url{https://tools.ssdc.asi.it/}} developed at the Space Science Data Center (SSDC). 

\begin{table*}
	\caption{Log of Swift XRT/UVOT observations used in our analysis}
	% title of Table
	\label{table:1}
	% is used to refer this table in the text
	\centering
	% used for centering table
	\begin{tabular}{c c c c c}
		% centered columns (4 columns)
		\hline\hline
		% inserts double horizontal lines
		Instrument & Observation ID & Start Time & XRT Exposure & UVOT Exposure \\
		&  & (MJD) & (ks) & (ks) \\
		
		% table heading
		\hline
		% inserts single horizontal line
		SWIFT-XRT/UVOT & 00031385003 & 59350.15 & 3.271 & 3.211\\
		% inserting body of the table
		SWIFT-XRT/UVOT & 00031385005 & 59352.48 & 1.221 & 1.152\\
		SWIFT-XRT/UVOT & 00031385006 & 59357.31 & 2.602 & 2.560\\
		SWIFT-XRT/UVOT & 00031385007 & 59360.35 & 2.225 & 2.134\\
		SWIFT-XRT/UVOT & 00031385008 & 59363.41 & 1.590 & 1.544\\
		SWIFT-XRT/UVOT & 00031385009 & 59365.21 & 1.334 & 1.293\\
		SWIFT-XRT/UVOT & 00031385011 & 59372.29 & 0.731 & 0.711\\
		
		\hline
	\end{tabular}
\end{table*}

\section{Results}
In this section, we describe our results obtained from the spectro-temporal analysis and we discuss the significance of the results in multi-waveband SED modelling. 

\subsection*{Light curve analysis-Gamma ray}
\begin{figure*}
	\centering
	\includegraphics[scale=0.4]{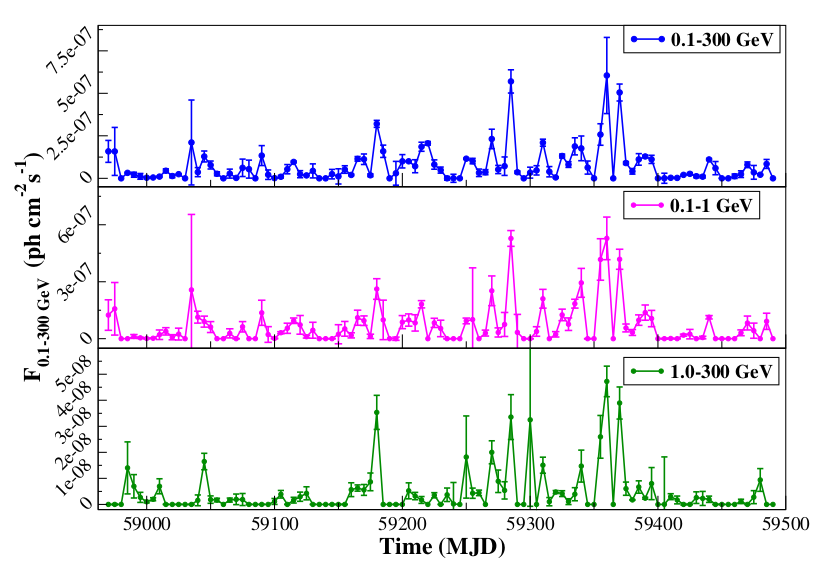}
	\caption{\textit{Top:} The full lightcurve for the source in 0.1-300 GeV energy corresponding to the period May 1, 2020 (MJD 58970) - September 30, 2021 (MJD 59500). There are several flaring episodes within this period. \textit{Middle:} Lightcurve for the 0.1-1 GeV band. We observe multiple flaring events here as well, the most prominent being around June 05, 2021 (MJD $\sim$ 59370). \textit{Bottom} Lightcurve for the 1-300 GeV band. We observe strong flaring activity only around June 15, 2021. }
	\label{fig1}
\end{figure*}

In Figure-\ref{fig1}, we provide the $\gamma$-ray lightcurve for the source in 0.1-300 GeV corresponding to the time between May 1, 2020 - September 30, 2021. In panels (b) and (c), we also show the same light curve for the lower energy band (LEB, 0.1-1 GeV) and higher energy band (HEB, 1-300 GeV). In Section 3.1, we describe the variability properties and the rise as well as decay times of the prominent flares that we calculated. We provide the correlation features between the LEB and HEB in Section 3.2, which hints at possible delays of onset of these flares in different wavebands and in turn can indicate the possible emission mechanisms. 

We observe that the entire domain of observation is interspersed with several flaring episodes, but frequent and short-duration flaring events are most prominent during the $\sim$ 150 day span from February 05, 2021 (MJD 59250) to July 05, 2021 (MJD 59400). 

\subsection{Estimation of variability}
In order to quantify the flux variability, we use the fractional variability amplitude \citep{Vaughan_2003} defined as
\begin{equation}
	F_{var} = \sqrt{\frac{\sigma_{XS}^2}{\mu^2}},
\end{equation}

\noindent
where $\sigma_{XS}$ is the excess variance of a time series with $x_i ~ \text{counrs s}^{-1}$ \citep{nandra1997asca} and $\mu$ is the sample mean. The excess variance is in turn defined as 

\begin{equation}
	\sigma_{XS}^2 = \sigma^2 - \frac{1}{N}\sum_{i}^N{\sigma_i}^2,
\end{equation}
where $\sigma$ is the standard deviation of the sample of size N and $\sigma_i$ is the statistical uncertainty of each of the data comprising the full sample. The normalized excess variance is defined to be $\sigma_{NXS} = {\sigma_{XS}^2}/{\mu^2}$, and the fractional variability happens to be the square root of this normalized excess variance. The error corresponding to the fractional variability is estimated as 

\begin{equation}
	err(F_{var}) = \sqrt{\Bigg( \sqrt{\frac{1}{2N}}.\frac{\sigma^2}{\mu^2F_{var}} \Bigg)^2 + \Bigg( \sqrt{\frac{\sigma^2}{N}} \frac{1}{\mu}\Bigg)^2},
\end{equation}
where the meaning of the respective quantities are as explained above \citep{prince2018fermi}. \par 

During the initial phase of low flaring activity (between MJD = 58970 and MJD = 59250), we observe the normalized excess variance in 0.1-300 GeV to be $0.85$, and the fractional variability to be $0.92\pm0.07$. However, towards the final later segment of observation (beyond MJD = 59250) where frequent flaring activities are noticed, the fractional variability becomes $1.18\pm0.70$.

\begin{figure*}
	\centering
	\includegraphics[scale=0.4]{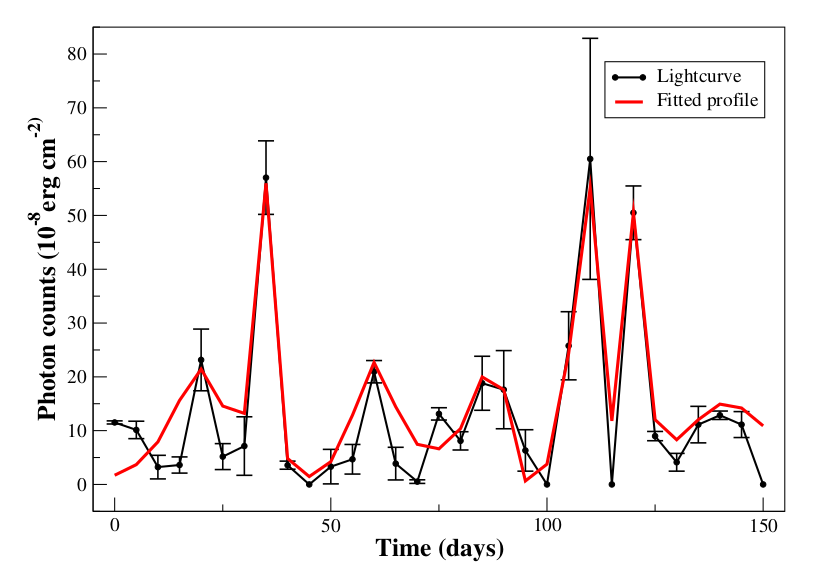}
	\caption{The flaring zone between MJD = 59250 and MJD = 59400 is fitted with sum of exponential profiles. The time-stamp in the X-axis is relative to MJD = 59250.}
	\label{fig2}
\end{figure*}

The rise and decay times of the flaring events were calculated by fitting the 150 days long $\gamma$-ray lightcurve with sum of exponential profiles of the form
\begin{equation}
	f(t) = F_0 + \sum_{i=1}^{N} 2F_i\Bigg(\exp\Bigg[{\frac{t_0-t_i}{T_r}\Bigg]} + \exp\Bigg[{\frac{t_i-t_0}{T_d}}\Bigg]\Bigg)^{-1},
\end{equation}
where $F_0$ is the baseline flux, $F_i$ is the source flux at time $t_0$  representing the approximate amplitude of the flare, $T_r$ and $T_d$ are respectively the rise and decay times of the flares. In HEB, we observe frequent flaring activity, but the photon intensity in this band is at least one order of magnitude lesser compared to broadband flux (0.1-300 GeV). We examine the rise and decay time, as well as the flux doubling time-scale only for the 150 days time window spanning between MJD = 59250 and MJD = 59400 by fitting the broadband flaring region using a combination of sum of exponential profiles, as shown in Figure-\ref{fig2}. Between MJD = 59265 and MJD = 59270, the 0.1-300 GeV flux rises from ($3.61 \pm 1.52$)$\times{10^{-8}} ~ \text{erg cm}^{-2}\text{s}^{-1}$ to ($2.31 \pm 0.57$)$\times{10^{-7}} ~ \text{erg cm}^{-2}\text{s}^{-1}$, and on MJD = 59275 the flux again drops to ($5.17 \pm 2.40$)$\times{10^{-8}} ~ \text{erg cm}^{-2}\text{s}^{-1}$. The computation of the flux doubling time was undertaken using the formula
\begin{equation}
	F(t_2) = F(t_1)2^{\frac{t_2 - t_1}{\tau_D}},
\end{equation}
where $F_{t1,t2}$ are the fluxes at time instances $t_1$ and $t_2$ respectively and $\tau_D$ is the flux doubling time. From this formula, we find the flux doubling time for this flare to be 3.22 days. The rise and decay time obtained from the sum of exponential fits were respectively found to be 2.40 days and 0.55 days respectively. Subsequently, between MJD = 59300 and MJD = 59350, we do not observe either a fully quiescent state or a full-fledged flare, and we call this a `plateau' state after \citep{prince2017long}. 

The average flux from the source is $(8.10\pm2.10)\times10^{-8}~ \text{erg cm}^{-2}\text{s}^{-1}$, with the flux peaking at $(1.78\pm0.17)\times10^{-7}~ \text{erg cm}^{-2}\text{s}^{-1}$. Subsequent to MJD = 59350, corresponding to which simultaneous X-ray data is available, we observe twin sharp flaring events within a span of $\sim$ 25 days. The peak flux of these individual flaring events were $(6.05\pm2.24)\times10^{-7}~ \text{erg cm}^{-2}\text{s}^{-1}$ and $(5.04\pm0.50)\times10^{-7}~ \text{erg cm}^{-2}\text{s}^{-1}$ respectively. The rise and decay times for the first twin flare appeared to be 1.6 days and 2.9 days respectively, and corresponding to the second twin flare they turned out to be 0.9 days and 1.3 days respectively. The shortest flux doubling time-scale thus turns out to be 0.9 days, which can be used to constrain the physical size of the emission region as well. From the shortest flux doubling time-scale of 0.9 days, we quantify the variability time-scale to be $t_{\text{var}} = \tau_D \times \ln{2} = 0.63$ days.

\begin{table*}
	\caption{Results from spectral analysis of \textit{Swift}-XRT/UVOT and \textit{Fermi}-LAT data}
	% title of Table
	\label{table:2}
	% is used to refer this table in the text
	\centering
	% used for centering table
	\begin{tabular}{c c c c}
		% centered columns (4 columns)
		\hline\hline
		% inserts double horizontal lines
		Instrument & Parameter & Value & Units \\
		
		% table heading
		\hline
		% inserts single horizontal line
		& Spectral index ($\alpha$) & $-2.17\pm0.03$ & - \\
		\textit{Fermi}-LAT & Prefactor ($N_0$) & $2.36\pm0.11$ & $10^{-11}~\text{ph cm}^{-2}\text{s}^{-1}\text{MeV}^{-1}$ \\
		& Flux ($F_{\text{0.1-300 GeV}}$) & $3.02\pm0.14$ & $10^{-7}~\text{ph cm}^{-2}\text{s}^{-1}$ \\
		\hline
		
		& $\Gamma_X$ & $1.58\pm0.05$ & - \\
		\textit{Swift}-XRT & $K$ & $3.74\pm0.17$ & $10^{-4}~\text{ph cm}^{-2}\text{s}^{-1}\text{keV}^{-1}$ \\
		& Flux ($F_{\text{0.3-8.0 keV}}$) & $2.43\pm0.11$ & $10^{-12}~\text{erg cm}^{-2}\text{s}^{-1}$ \\
		\hline
		& u band flux & $4.22\pm0.10$ & $10^{-12}~\text{erg cm}^{-2}\text{s}^{-1}$ \\
		& b band flux & $6.50\pm0.14$ & $10^{-12}~\text{erg cm}^{-2}\text{s}^{-1}$ \\
		\textit{Swift}-UVOT & v band flux & $8.15\pm0.22$ & $10^{-12}~\text{erg cm}^{-2}\text{s}^{-1}$ \\
		& w1 band flux & $2.15\pm0.06$ & $10^{-12}~\text{erg cm}^{-2}\text{s}^{-1}$ \\
		& w2 band flux & $1.03\pm0.03$ & $10^{-12}~\text{erg cm}^{-2}\text{s}^{-1}$ \\
		& m2 band flux & $1.57\pm0.04$ & $10^{-12}~\text{erg cm}^{-2}\text{s}^{-1}$ \\
		
		% inserting body of the table

		\hline
	\end{tabular}
\end{table*}

\subsection{Timing Analysis}
\begin{figure}
	\centering
	\begin{subfigure}[b]{0.45\textwidth}
		\centering
		\includegraphics[width=\textwidth]{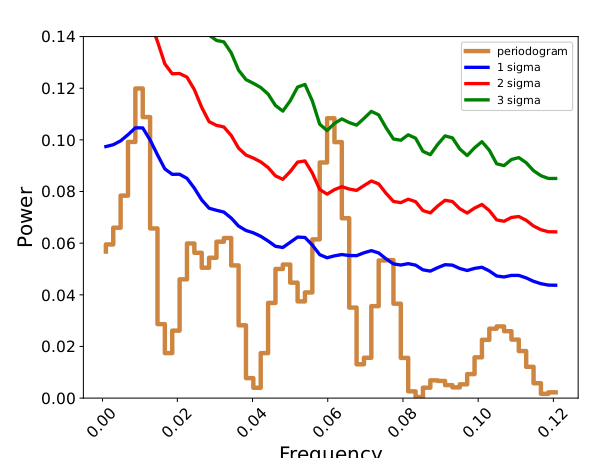}
		\caption{Low frequency periodogram}
		\label{periodogram_low}
	\end{subfigure}
	\hfill
	\begin{subfigure}[b]{0.45\textwidth}
		\centering
		\includegraphics[width=\textwidth]{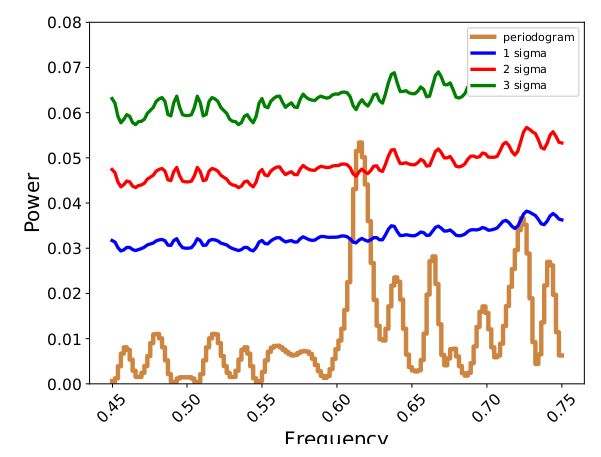}
		\caption{High frequency periodogram}
		\label{periodogram_high}
	\end{subfigure}
	\hfill
	\caption{Low and high frquency periodograms. In low frequency, we observe a $>3\sigma$ significant peak at $\sim$ 17 days, while in high frequency the peak is only $\sim 2.5\sigma$ significant.}
	\label{periodograms}
\end{figure}

We have carried out a temporal analysis of the $\gamma-$ ray emission from a blazar TXS 1700+685 with the \textit{Fermi-} LAT data from May 1, 2020 to September 30, 2021. In the low-frequency domain, we have detected a sharp peak in the periodogram at $\sim 0.06~ \text{days}^{-1}$ which has $\sim 3\sigma$ significance (Figure-\ref{periodogram_low}). The significance of the periodogram peak is estimated by the method proposed by \citep{10.1093/mnras/stt764} and simulating 1000 $\gamma$-ray light curve \footnote{https://github.com/samconnolly/DELightcurveSimulation}, as elucidated in \citep{2015arXiv150306676C}. However, we also detect another peak at $\sim 0.62~ \text{days}^{-1}$ which has marginally $> 2\sigma$ detection significance (Figure-\ref{periodogram_high}).   

A variety of mechanisms have been proposed to explain the background physical behavior in blazer $\gamma$-ray emission. These include periodically varying Doppler beaming from precessing jet \citep{marscher1985models,camenzind1992lighthouse,abraham2000precession,caproni2013parsec}, processed related to the innermost stable orbit of the accretion disc \citep{broderick2006imaging,pihajoki2013short} and several others.

One possible explanation of the QPO of a blazar involves an accretion disc hot-spot orbiting close to the innermost stable circular orbit of the central SMBH \citep{gupta2019mnras}. These hotspot emissions are quasi-thermal and could be directly observed in the optical/X-ray band. The optical/X-ray emission could generate the variation in seed photons and after external Compton interaction with jet, it can trigger the flux modulation in the $\gamma-$ray lightcurve \citep{gupta2017peculiar}. However, in this case, the expected period in the optical/X-ray band of QPO depends on the mass and spin of the central SMBH. Furthermore, it would not be the same in the gamma-ray band, which would be Doppler boosted. Due to the lack of optical/X-ray data, we can not comment on optical/X-ray variability. \par

%A third possibility is the precession of the jet of a blazar. These types of models could produce QPOs in the light curve of a blazar in the time scale $\sim$ 1 year or more \citep{rieger2004geometrical}), which does not agree with the present observations.  \par 

Another potential explanation of the QPO in Gamma-ray energy is instability in the disk/jet system. Magnetically choked accretion flows (MCAFs) in general-relativistic three-dimensional magnetohydrodynamic simulations are observed to produce quasi-periodic fluctuations in the energy outflow efficiency of the relativistic jet close to the black hole \citep{tchekhovskoy2011efficient}. The dominant mode of these fluctuations has a period of $\sim 70r_g/c$ for the spin $a=0.9375$. For a central mass of $\sim10^8$ M$_\odot$ has a period of $\sim1$d when the Schwarzschild radius $r_s \approx 3\times 10^{11}$m. Slower-spinning black holes are expected to have longer periods \citet{McKinney2021.423.3083M}. It is unclear, whether these oscillations can produce an observable signature in the radio jet or not. However, The different time-scales make it improbable that this mechanism may explain the QPO seen in the Gamma-ray light curve in TXS 1700+685. \par 

The quasi periodic flux variation can be induced because of the shock advancement along the internal helical structure of the jet, or the precession or twisting configuration of the jet \citep{camenzind1992lighthouse, mohan2015kinematics}. There could be several underlying mechanisms behind a precessing jet. The modulation of the jet axis could be induced by the Lense–Thirring precession of the disk \citep{fragile2009general}. It can also be triggered if the AGN is part of a binary supermassive black hole system \citep{Ackermann_2015,sandrinelli2016gamma, sandrinelli2016quasi}. However, the oscillation period arising out of such a mechanism appears to be $\sim$ few years, which has been detected recently in the case of a few AGN sources \citep{Ackermann_2015,sandrinelli2016gamma, sandrinelli2016quasi}. In our case, we found a periodicity about $\sim$ 17 days, which is much lower than the expected value in the case of binary SMBH scenario. \par 

However, in the scenario of the helical motion of the plasma blob along the internal helical structure of the jet \citet{mohan2015kinematics}, the variation of the Doppler boosting factor in association with the viewing angle of the emission region can induce a significant flux modulation in the high energy domain. Depending on the viewing angle and the boosting factor, the expected QPO may vary from day to month time scale. Our analysis of the gamma-ray emission from TXS 1700+685 suggests a similar time scale expected from this model. The observed modulation of the flux could be the enhanced emission region moving helically within the curved-jet or curved helical jet itself \citet{sarkar2021multiwaveband}. As we have a limited idea about the emission mechanism of blazar and lack of multiwavelength data of this source, it is possible that the observed modulation in gamma-ray flux is due to some other intrinsic effect or combinations of these effects. 

\subsection{Auto Correlation Function}
\begin{figure}
		\centering
		\includegraphics[scale=0.3]{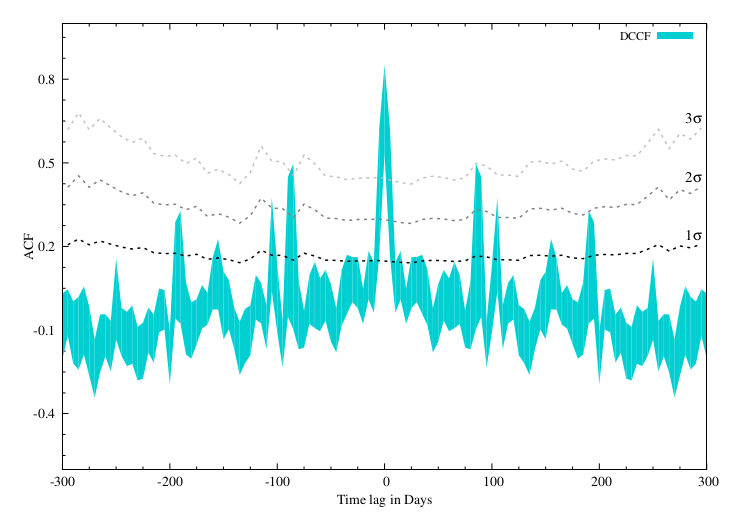}
		\caption{ We have also calculated the confidence of the peaks in correlation function (DCF). Here we found the peak near $\pm 82$ days have a confident upto $3\sigma$. }
		\label{fig3}
	\end{figure}

We have performed Auto Correlation Function (ACF) for the $\gamma$-ray light curve (0.3 to 500 GeV) using the discrete correlation function (DCF)\footnote{https://github.com/astronomerdamo/pydcf/blob/master/dcf.py} \citep{1988ApJ...333..646E} and $\zeta$-discrete correlation function ($\zeta$-DCF)\footnote{https://www.weizmann.ac.il/particle/tal/research-activities/software} \citep{2014ascl.soft04002A} for comparison. Both correlation functions are followed the same pattern. The most correlated peak is found at zero time lag which is expected as
here we have done the ACF analysis. The 2nd most significant peak is found $82\pm5$ days on both sides of the main peak. The correlation coefficient value of this peak is about 50 for both peaks. 
In estimating the significance of the ACF peak, the same procedure as elucidated in the previous section has been followed.
We perform the computation of ACF over 1000 $\gamma$-ray light curves and estimated the 1$\sigma$, 2$\sigma$ and 3$\sigma$ for each time lag. The  $\sigma$ levels (black and gray curves) are plotted on the correlation function as shown in Figure-\ref{fig3}. We have found 82$\pm$5 days peaks are significant up to 3$\sigma$. 

% \begin{figure*}[h!]
	%	\centering
	%	\includegraphics[scale=1]{figures/dcf.eps}
	%	\caption{\textit{Top:(a)} The variation of high energy $\gamma$-ray (0.3 - 500 GeV) count with respect to time in days with 5 day binning is shown. \textit{bottom:(b)} The auto correlation functions which are calculated using DCF and $\zeta$-DCF algorithm, for the light curve is shown. \textit{bottom:(c)} Periodogram is also calculated using Lomb-Scargle method for the high energy light curve. We have also calculated the Power Spectral Density (PSD) for the given light curve.  }
	%	\label{fig1}
	%\end{figure*}

	\subsection{Examination of high energy photons}
\begin{figure*}
		\centering
		\includegraphics[scale=0.4]{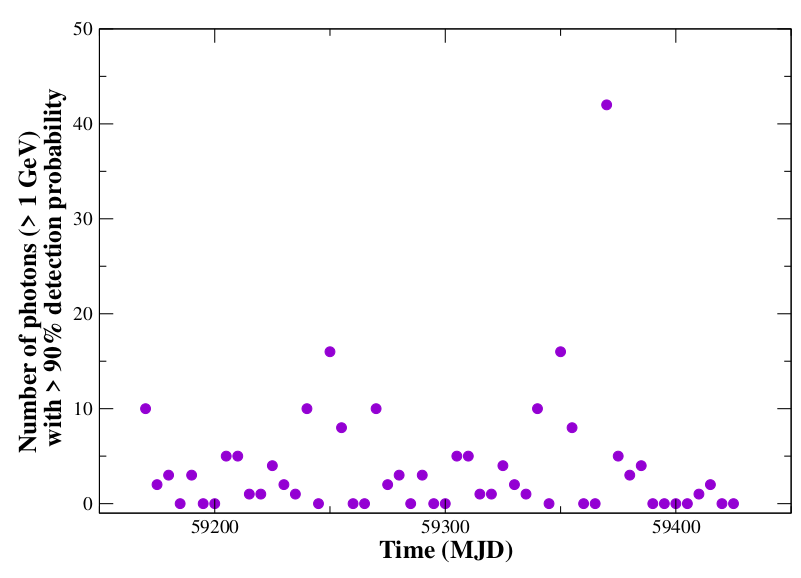}
		\caption{ Time variation of the number of high energy ($> 1$ GeV) photons with $>90\%$ probability of detection. We detect this number to be peaking on MJD = 59370.}
		\label{fig4}
	\end{figure*}

	In order to estimate the dominance of high energy photons during the flaring region, we carried out the \textit{Fermi}-LAT analysis with $2^{\circ}$ of ROI. We choose this threshold because the flux of the source was low, the likelihood analysis could not be performed for most of the time intervals with smaller ROI. We do not perform the analysis with the ULTRACLEAN class of events because of the same reason. In Figure-\ref{fig4}, we plot the lightcurve corresponding to the high energy ($> 1$ GeV) photons detected within $2^{\circ}$ ROI within the flaring phase. In (Figure-\ref{fig4}), we show the number of such photons with $> 90\%$ probability of detection at successive time intervals. \par 
	
	We observe the highest number of high energy photons (total 42) on MJD = 59370, and the highest energy of these photons being 6.74 GeV with the detection probability 99.80\%. The number of such high-energy photons during the pre-flare plateau region is significantly lower, with the highest number being 10 on MJD = 59270. The highest energy of the detected photons with high detection significance corresponding to the entire span of observation pertaining to our analysis, however, turns out to be 25.41 GeV on MJD = 59275 with a detection probability of 99.97\%. Such high energy photons could be
	produced as a consequence of the external Compton scattering of the photons originating from the broad-line region (BLR), or dusty torus by the jets or synchrotron self-Compton emission. 
	
	\subsection{Gamma-ray Spectral energy distribution during flaring phase}
\begin{figure*}
		\centering
		\includegraphics[scale=0.4]{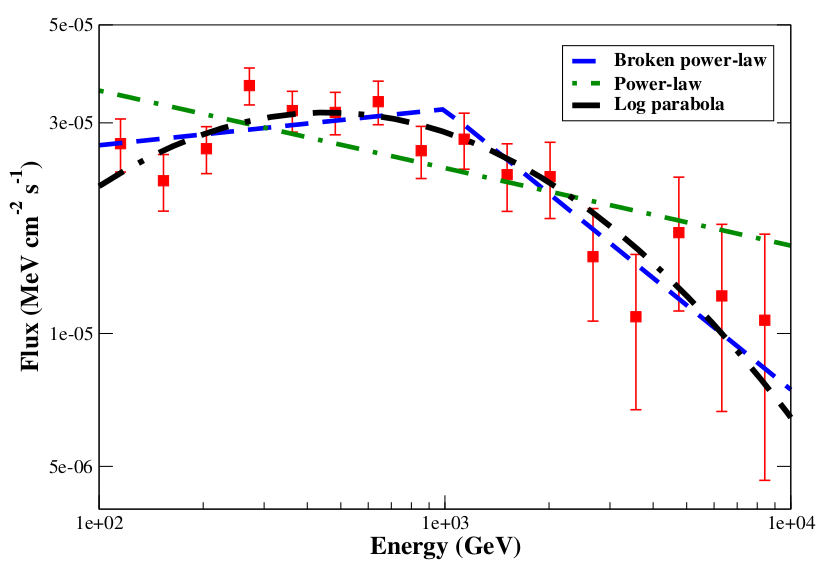}
		\caption{SED during the flaring phase approximated using different profiles (power-law, broken power-law, and log-parabola). We observe a statistically significant break at $\sim$ 1 GeV.}
		\label{fig5}
	\end{figure*}

	In this section, we describe the key features regarding the $\gamma$-ray spectral energy distribution (SED). We attempt to approximate the shape of the SED using different profiles: power-law (PL), log-parabola (LP), and broken power-law (BPL) as defined in Equations 1-3. The choice of these functional forms is motivated by the spectral fitting over equispaced logarithmic energy bins of blazar flares in 100 MeV - 200 GeV range \citep{ackermann2010fermi}. In order to determine the quality of the corresponding spectral fits, the Log(likelihood) values obtained from the unbinned likelihood analysis have also been examined. In Figure-\ref{fig5}, we show the comparison of different profiles. The spectral index corresponding to the power-law fit turns out to be $2.18\pm0.03$. However, even by visual inspection the presence of spectral curvature is detected, which motivates us to examine the alternative phenomenological models such as logparabola (LP) and broken power law(BPL). The significance of the spectral curvature is determined via $\text{TS}_{\text{curve}}$, which is defined as the $\text{TS}_{\text{curve}} = 2(\mathcal{L}(BPL/LP) - \mathcal{L}(PL))$. The spectral curvature is significant only if $\text{TS}_{\text{curve}} > 16$ \citep{nolan2012fermi}, where L represents the likelihood of the function. In both of the cases (LP and BPL) we found this criterion to be satisfied, implying that spectral curvature is statistically significant. In the case of LP, we find $\alpha = 2.26\pm0.05$ and $\beta = 0.17\pm0.04$, with the $TS_\text{curve} = 18$. On the other hand in the case of BPL fit, we find the spectral index before the break at 1 GeV to be $1.92 \pm 0.07$ and after the break to be $2.63 \pm 0.11$, with $TS_\text{curve} = 34$. Thus, The BPL photon index demonstrates the trend of being $< 2$ before the break, and after $> 2$ after the break. Such a falling spectrum has earlier been observed in the case of several other FSRQs as well, 3C 454.3 \citep{2011ApJ...733L..26A}, 3C 279 \citep{paliya2015violent,prince2020broadband}. \par 
	
	These results can also serve to constrain the possible origin of the $\gamma$-ray emission region. If the power-law could approximate the SED well, it would imply that the emission region is outside the broad-line region (BLR). However, the presence of a significant spectral curvature implies that the photons are further reprocessed in the BLR (like photon-photon pair production), which attenuates the $\gamma$-ray flux leading to a spectral break \citep{liu2006absorption}. The presence of the significant cut-off at 1 GeV explains the non-detection of very high energy photons throughout the period of observation, and also implies that the 25.41 GeV photon detected on MJD = 59275 might have originated from outside of the BLR. 
	
	\subsection{Broadband SED modelling}
	Modeling the multi-waveband SED turns out to be a powerful tool in understanding the flaring mechanism associated with the source under consideration. In our work, we consider a one-zone leptonic model of jet acceleration based on the stochastic
	or shock acceleration mechanism \citep{ghisellini2009canonical,2011ApJ...739...66T}. We use the publicly available \textsc{jetset} implementation for the purpose of our SED modelling \citep{2006A&A...448..861M,2009A&A...501..879T,2011ApJ...739...66T}. In this model, it is assumed that a spherically symmetric blob of size $R$ filled with relativistic population of electrons is moving through an entangled magnetic field ($B$). The leptons in the jet are accelerated to an ultra-relativistic regime via a stochastic or shock acceleration mechanism inside the jet. The plasma blobs are assumed to be moving at an angle $\theta$ relative to the observer's line of sight, implying that the emission will be beamed by a factor $\delta = 1/[\Gamma(1 - \beta\cos{\theta})]$, where $\Gamma$ is the bulk Lorentz factor of the blob. We assume the relativistic population of electrons within the blob is satisfying an empirical broken power-law distribution of the form,
	
	\begin{equation}
		n({\gamma}) =
		\begin{cases}
			K_1\gamma^{p_1} & \text{when $\gamma_{\text{min}} \le \gamma \le \gamma_{\text{break}}$}\\
			K_2\gamma^{p_2} & \text{when $\gamma_{\text{break}} \le \gamma \le \gamma_{\text{max}}$}\\
			%0 & \text{otherwise}
		\end{cases}       
	\end{equation}
	\noindent
	where $\gamma_{min}, \gamma_{max}$ and $\gamma_{\text{break}}$ are respectively the minimum, maximum, and the break energies of the relativistic electrons, and $p_1$ and $p_2$ are the spectral indices below and above the break. The normalization constants $K_1$ and $K_2$ are related to each other by $K_2 = K_1(\gamma_{\text{break}})^{p_2 - p_1}$. \par 
	
	\begin{figure*}
		\centering
		\includegraphics[scale=0.4]{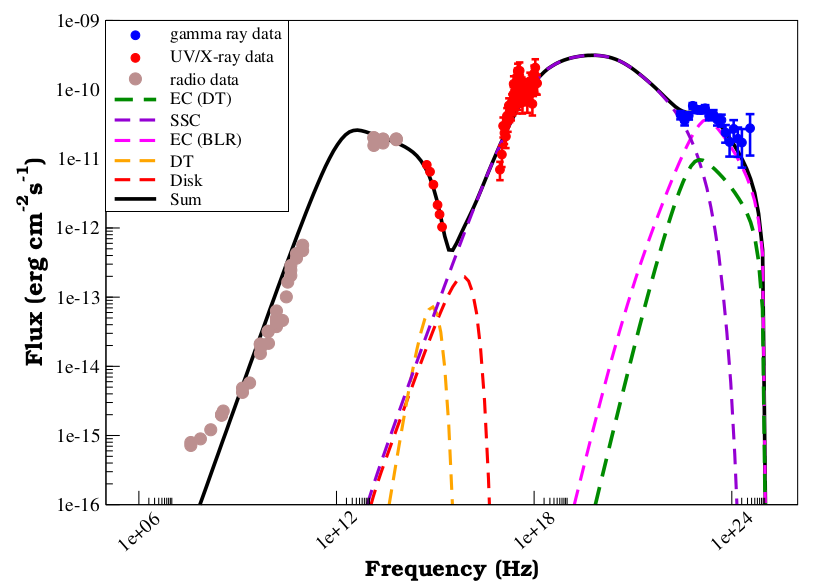}
		\caption{Multiwavelength SED suring the flaring phase modeled in one-zone leptonic framework. The best-fit model parameters are listed in Table 3.}
		\label{fig6}
	\end{figure*}

	The first hump of the broadband SED is produced by the synchrotron emission as a consequence of the interaction of relativistic electrons with the magnetic field. These synchrotron photons serve as the seed photons for the inverse Compton process by the same population of relativistic electrons named as synchrotron self-Compton (SSC), and this SSC leads to the high energy hump. Further, several other processes which fall in the category of external Compton (EC) process also contribute to the genesis of this hump. The seed photons for the EC emission come from \\
	(i) direct emission from the accretion disk, \\
	(ii) reprocessed emission from the infrared region of the surrounding dusty torus (DT), \\
	(iii) reprocessed emission in the optical/UV domain from the BLR. \\
	
	We consider the seed photons from BLR, DT as well as the disk for the purpose of our broadband SED modeling. The thermal emission of the accretion disc is approximated by the multi-temperature blackbody radiation, such that the radial profile of temperature is provided as 
	\begin{equation}
		T^4(r) = \frac{3R_SR_D}{16\epsilon\pi\sigma r^3}\Bigg( 1 - \sqrt{\frac{3R_S}{r}} \Bigg) K,
	\end{equation}
	where $R_S$ is the Schwarzschild radius, $\epsilon$ is the efficiency of accretion, $\sigma$ is the Stefan-Boltzmann constant. The outer boundary of the disc is assumed to be $500R_S$. In our SED modeling, we let the parameters regarding the electron distribution as well as the magnetic field vary, while the inner and outer extent of the BLR ($1.0\times10^{18}$ cm and $2.0\times10^{18}$ cm respectively) as well as the size of the DT ($5.0\times10^{18}$ cm) are left frozen. The broadband spectral fitting is shown in Figure-\ref{fig6} and the fitted parameters are provided in Table 3.
	
	\begin{table}
		\caption{\label{t7}Broadband SED modeling parameters. }
		\centering
		\begin{tabular}{lcc}
			\hline\hline
			Parameter & Unit & Value\\
			\hline
			$B^{1}$ & Gauss & 0.12  \\
			&&\\
			$\theta^{2}$ & degree & 3.60  \\
			&&\\
			$\Gamma^{3}$ & - & 7.08  \\
			&&\\
			$\delta^{4}$ & - & 11.78  \\
			&&\\
			$\gamma_{\text{min}}^{5}$ & - & 4.87$\times10^{2}$  \\
			&&\\
			$\gamma_{\text{max}}^{6}$ & - & 1.05$\times10^{4}$   \\
			&&\\
			$\gamma_{\text{break}}^{7}$ & - & 4.19$\times10^{3}$  \\
			&&\\
			$\alpha_1^{8}$ & - & 3.32  \\
			&&\\
			$\alpha_2^{9}$ & - & 3.15  \\
			&&\\
			$U_e^{'10}$ & $\text{erg cm}^{-3}$ & 0.51\\
			&&\\
			$U_B^{'11}$ & $\text{erg cm}^{-3}$ & $6.09\times10^{-4}$\\
			&&\\
			$U_e^{'}/U_B^{'12}$ & - & 839.08\\
			&&\\
			$N^{13}$ & $\text{cm}^{-3}$ & $7.41\times10^2$ \\
			\hline
			& \textbf{Parameters external to the jet} & \\
			\hline
			$L_d^{14}$ & $10^{45}~\text{erg s}^{-1}$ & 1.73 \\
			&&\\
			$T_{DT}^{15}$ & K & 1.28$\times10^4$ \\
			&&\\
			$\tau_{DT}^{16}$ & - & 0.93 \\
			&&\\
			$R_{DT}^{17}$ & $10^{18}\text{cm}$ & 5.0 \\
			&&\\
			$R_{BLR}^{\text{in}~ 18}$ & $10^{18}\text{cm}$ & 1.0 \\
			&&\\
			$R_{BLR}^{\text{out}~ 19}$ & $10^{18}\text{cm}$ & 2.0 \\
			&&\\
			$\tau_{BLR}^{20}$ & - & 0.24 \\
			
			\hline
		\end{tabular}
		\footnotesize{$^1$ Magnetic field, $^2$ Viewing angle relative to the line of sight, $^3$ Bulk Lorentz factor of the jet, $^4$ Doppler beaming factor, $^5$ Minimum Lorentz factor of the injected electrons, $^6$ Maximum Lorentz factor of the injected electrons, $^7$ reak Lorentz factor of the injected electrons, $^8$ Low energy spectral index, $^9$ High energy spectral index, $^{10}$ Electron energy density in comoving frame, $^{11}$ Magnetic field energy density in comoving frame, $^{12}$ Equipartition value, $^{13}$ Particle density, $^{14}$ Disk luminosity, $^{15}$ Temperature of the dusty torus, $^{16}$ Optical depth of the dusty torus, $^{17}$ Distance of the dusty torus, $^{18}$ Inner radius of the BLR, $^{19}$ Outer radius of the BLR, $^{20}$ Optical depth of the BLR,}\\
			%\tablefoot{ \tablefoottext{1}{ Magnetic field.}, \tablefoottext{2}{ Viewing angle relative to the line of sight}, \tablefoottext{3}{ Bulk Lorentz factor of the jet}, \tablefoottext{4}{ Doppler beaming factor},  \tablefoottext{5}{ Minimum Lorentz factor of the injected electrons},  \tablefoottext{6}{ Maximum Lorentz factor of the injected electrons},  \tablefoottext{7}{ Break Lorentz factor of the injected electrons}, \tablefoottext{8}{ Low energy spectral index}, \tablefoottext{9}{ High energy spectral index}, \tablefoottext{10}{ Electron energy density in comoving frame}, \tablefoottext{11}{ Magnetic field energy density in comoving frame}, \tablefoottext{12}{ Equipartition value}, \tablefoottext{13}{ Particle density},\tablefoottext{14}{ Disk luminosity}, \tablefoottext{15}{ Temperature of the dusty torus}, \tablefoottext{16}{ Optical depth of the dusty torus}, \tablefoottext{17}{ Distance of the dusty torus},\tablefoottext{18}{ Inner radius of the BLR},\tablefoottext{19}{ Outer radius of the BLR},\tablefoottext{20}{ Optical of the BLR}.}
	\end{table}
	\section{Results and Discussions}
	In this section, we discuss the implications of our timing and correlation study, as well as the broadband spectral fitting.
	
	\subsection{Self-correlation and the plausible effect of gravitational lensing}
	
	The gravitational light-bending effect of the emitted radiation of quasars by the intervening foreground sources has been well recognized in the literature \citep{torres2003gravitational}. For a gravitationally lensed blazar source, the $\gamma$-ray lightcurve is expected to show some time-delay, as has already been found in the case of PKS 1830-211 \citep{barnacka2011first} and B0218+357 \citep{cheung2014fermi}. The resolution of \textit{Fermi}-LAT is not sufficient to resolve the lensed images located within 1 arc second spatial resolution \citep{rao19881830}, but the time-lag between different photon paths detected from the self-correlation of the $\gamma$-ray lightcurves could be a plausible way to detect such lensing effects \citep{abdo2015gamma,abhir2021study}.\par
	In our case, we have detected a strong self-correlation ($>3\sigma$ detection significance) of the 0.1-300 GeV $\gamma$-ray lightcurve at a time-delay of 82 days (Figure-\ref{fig3}). We conjecture the presence of such a strong signature of a time delay to be a consequence of the microlensing effect of the $\gamma$-ray photons originating from different emission regions. However, conclusive pieces of evidence can only be gathered from detailed radio monitoring of this source, since it would then be possible to infer the presence and impact of such lensing effects using flux ratios \citep{lovell1996pks}. Also, since the emission region corresponding to radio and $\gamma$-ray could be different, emission size induced chromaticity would be an important indicator to probe such effects.

	\subsection{Estimation of minimum Doppler factor}
	The numerical value corresponding to the minimum Doppler factor can be estimated from the highest energy photon detected during the observation. Assuming the optical depth $\tau_{\gamma\gamma}(E_h)$ of the highest energy photon $E_h$ to $\gamma\gamma$ interaction being 1, the expression for the minimum Doppler factor could be read as
	\begin{equation}
		\delta_{\text{min}} = \Bigg ( \frac{\sigma_T D_L^2 (1+z)^2 f_e E_h}{4t_\text{var}m_ec^4} \Bigg)^{1/6},
	\end{equation}
	\noindent
	where $\sigma_T$ is the Thomson cross-section, $D_L$ is the luminosity distance of the source which turns out to be 1547 Mpc with under standard cosmology parameters, $z (= 0.3)$  is the redshift of the source, $f_e$ is the 0.3-8.0 keV X-ray flux, $t_\text{var}$ is the observed variability time-scale of the source, $E_h$ is the energy of the highest energy photon. In our case, during the flare we observe the fastest variability time-scale to be 0.63 days, and the highest-energy photon with significant detection probability to be 6.67 GeV. With these parameters, $\delta_{\text{min}}$ turns out to be 8.76. Our fitted estimation of the Doppler factor (=11.78) turns out to be more than this threshold.
	
	\subsection{Broadband emission during the flaring state}
	In the broadband SED, each segment indicates some constraints on the allowed parameter space. In the case of blazars, some important constraints are:
	\begin{enumerate}
		\item The synchrotron emission component suffers a sharp decline below $10^{12}$ Hz because of the synchrotron self-absorption mechanism, in which the photons interact with the ambient electrons and lose energy. Since the radio data is above the predicted curve and demonstrates only the emission from the extended segment from the jet, radio data provides only a weak constraint on the low energy synchrotron spectrum.
		
		\item The optical/UV data is primarily contributed by the synchrotron and the optical emission from the disk, this segment of the SED provides constraints on the magnetic field ($B$), particle density ($N$) and the first spectral index ($\alpha_1$). 
		
		\item The high energy peak contributed by the X-ray/$\gamma$-ray is contributed by the synchrotron self-Compton (SSC) and external Compton (EC) processes. This segment is determined by the second spectral slope ($\alpha_2$), the physical distances of BLR and DT ($R_{\text{BLR}}$ and $R_{\text{DT}}$). 
		
		\item Overall emission is also a function of the bulk Lorentz factor of the jet ($\Gamma$), viewing angle ($\theta$), Doppler beaming factor ($\delta$).
	\end{enumerate}
	
	Because of the weak constraint provided by the radio data and given the fact that in the absence of simultaneous radio observation we are using only historical data for referencing purposes, the radio segment of the SED is not well approximated by our model, which is common in Blazar SEDs ( (e.g., \citet{10.1111/j.1365-2966.2012.21707.x}). The fitted spectrum is provided in (Figure-\ref{fig4}) and the fitted parameters are listed in Table 3. The lower energy hump is mostly contributed by the synchrotron component, and the contribution from the disk emission is found to be sub-dominant. Reharding the high energy hump, we observe the X-ray segment is well modeled with the SSC component, while the $\gamma$-ray segment is produced by both SSC and EC components. The generation of keV photons in synchrotron emission requires the presence of a strong magnetic field or the injection of a large number of high-energy electron populations. The equipartition value that we have observed in our case implies that the bulk of the power is carried away by the electron population, and the magnetic field carries only a tiny fraction of power. This is reflected from the low magnetic field ($\sim 0.12$ G) we have obtained from our spectral fit. Our estimation of the magnetic field is in agreement with the previous estimation for which the equipartition value is similarly on the higher side (e.g. in PKS 1424-418; \citet{10.1093/mnras/staa3639} for which the estimated magnetic field turns out to be $\sim 0.2$ G), and is at variance with low equipartition value estimation corresponding to stronger magnetic fields \citep{2014A&A...569A..40B}. The bulk Lorentz factor ($\Gamma$) and the Doppler factor ($\delta$) are also significantly lower compared to what was obtained in \citet{2014A&A...569A..40B}. \par 
	
	The highest-energy photon obtained during the flaring phase was found to be $6.67$ GeV. This is below the threshold for which the photons become opaque (more than $\sim$ 20 GeV) and are reprocessed by the BLR region. Because of this reason, we included the photons from DT, BLR as well as the disk as seeds for the EC mechanism. We observe the EC contribution by the BLR region to be dominant in the $\gamma$-ray segment, and the EC from the DT to be subdominant. The EC component corresponding to the disk photons seems to be negligible (turns out to be 4-5 orders of magnitude smaller compared to the other two components). \par 
	
	\subsection{Estimation of jet power}
	We have estimated the approximate power carried by different components of the jet, as well as the entire jet by the following equation
	
	\begin{equation}
		P_{\text{jet}} = \pi R_e^2 \Gamma^2 c(U_e^{\prime} + U_B^{\prime} + U_p^{\prime}),
	\end{equation}
	\noindent
	where the quantities $U_e^{\prime}, U_B^{\prime}$  and $U_p^{\prime}$ are respectively the energy densities corresponding to the electrons, magnetic field and protons in the co-moving frame of the jet (the un-primed quantities are in observer's rest frame while the primed quantities are in co-moving frame). The power contributed by the leptonic component is represented by
	\begin{equation}
		P_e^{\prime} = \frac{3\Gamma^2c}{4R_e}\int_{E_{\text{min}}}^{E_{\text{max}}} EQ(E) dE,
	\end{equation}
	\noindent
	where $Q(E)$ is the injected particle spectrum and $E_{\text{min}}$ and $E_{\text{max}}$ are the minimum and maximum energies of the electron calculated by multiplying the rest mass energy of the electron (in MeV) with $\gamma_{\text{min}}$ and $\gamma_{\text{max}}$ respectively. In order to calculate the power carried by the protons, we assume the ratio of the number of electron-positron pairs to the number of protons by 20:1. The power contributed by the magnetic field, on the other hand, can be calculated using
	
	\begin{equation}
		P_B^{\prime} = R_e^2 \Gamma^2 c \frac{B^2}{8}.
	\end{equation}
	\noindent
	where $B$ is the strength of the magnetic field obtained from the broadband fit. \par 
	
	Using the fitted parameters, we find the jet power contributed by the electrons and protons to be $1.61\times10^{43}~\text{erg s}^{-1}$ and $1.06\times10^{42}~\text{erg s}^{-1}$ respectively. The power deposited by the magnetic field component turns out to be $3.02\times10^{42}~\text{erg s}^{-1}$, which is subdominant compared to the leptonic component. The total jet power turns out to be $2.03\times10^{43}~\text{erg s}^{-1}$.
	
	\begin{figure}
		\centering
		\includegraphics[scale=0.4]{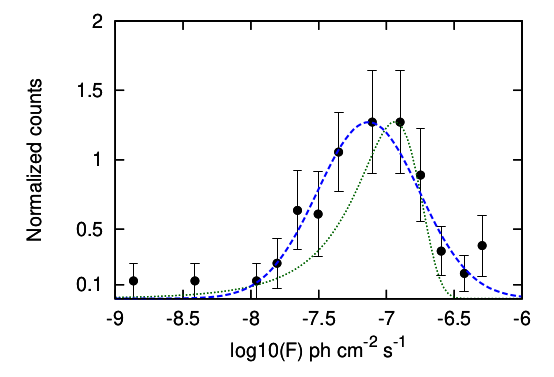}
		\caption{Histogram of the logarithm of three days binned gamma-ray flux. The green dotted line and the blue dashed line represent the Gaussian and lognormal PDFs, respectively.}
		\label{fig7}
	\end{figure}
	
	\begin{figure}
		\centering
		\includegraphics[scale=0.5]{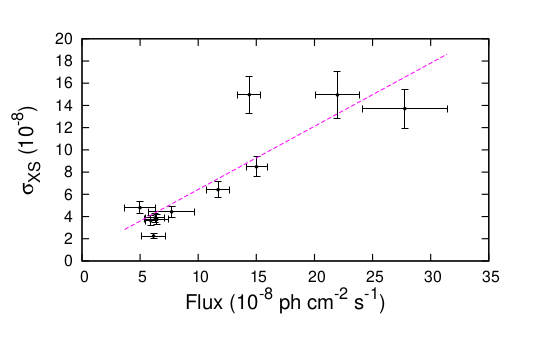}
		\caption{The rms-flux scatter plot, with a best fit line (magenta).}
		\label{fig8}
	\end{figure}

	\subsection{Flux distribution of TXS 1700+685}
	The flux distribution study can provide insight into the underlying mechanisms producing the variations in blazars. 
	The flux variability in blazars has been observed to be stochastic in nature. Typically, the flux distribution is Gaussian in the case of linear stochastic process, while a lognormal flux distribution (i.e. a Gaussian distribution in the logarithm of flux) is expected 
	for a non-linear stochastic process. Several blazars show lognormal flux distributions using multi-wavelength lightcurves, and at different time-scales \citep{2009A&A...503..797G, 2010A&A...520A..83H, 2010A&A...524A..48T,pankaj_ln, 2018MNRAS.480L.116S, 2020MNRAS.491.1934K,2020ApJ...895...90S, 2021MNRAS.502.5245P}. 
	The advent of high sensitive {\it Fermi}-LAT (100 MeV - 300 GeV) enables us to characterize the flux distribution study in $\gamma$-ray energy. Several blazars seen by {\it Fermi}-LAT, show lognormal behavior in their long-term $\gamma$-ray lightcurves \citep{Ackermann_2015, 2018RAA....18..141S, 2018Galax...6..135R, 2020ApJ...891..120B}. 
	The lognormal variability in blazar is usually explained by assuming the jet emission arising from non-linear multiplicative processes originated in the accretion disc. 
	However, minute-scale variability as observed in many blazars may not be produced by the fluctuations in the disc, and that favors the variability to produce in the jet itself.  
	The ``mini jets-in-a-jet" statistical model provides a possible explanation where the addition of emission from a large collection of mini jets can explain the lognormal flux distribution \citep{minijet}. Furthermore, \citealp{2018MNRAS.480L.116S} showed that the lognormal flux distribution can be obtained when the underlying particle acceleration process is time-dependent. In such cases, a Gaussian perturbation in the particle acceleration time-scale can give rise to a lognormal flux distribution and linear flux-rms correlations at high energies.
	
	In this work, we presented the flux distribution study for the source TXS 1700+685, with three days binned gamma-ray flux lightcurve. To select the statistically significant lightcurve, we considered the flux points for which flux is greater than 2-$\sigma$ level, such that ${\frac{F}{\Delta{F}}}$ > 2. 
	We produced the normalized histogram of the logarithm of flux and fit it with the Gaussian and lognormal probability density functions (PDFs, \citet{2018RAA....18..141S}) (Figure-\ref{fig7}). It is found that the lognormal function fits the distribution significantly better with reduced chi-square (${\chi^{2}}_{red}) {\approx}$ 0.91 for 11 degrees of freedom (dof), than the Gaussian function (${\chi^{2}}_{red} {\approx}$ 2.29 for 11 dof). The linear dependence of the mean flux $<F>$ with its excess root mean square (rms) variability, is an essential feature for the lognormal process. We estimated the excess variance \citep{Vaughan_2003}, $\sigma_{XS}= \sqrt{\sigma^2-\overline{\sigma_{err}^2}}$ with $\sigma^{2}$ and $\overline{\sigma_{err}^2}$ represent the variance and the average of the square of the measurement errors, using 50d binned data. The rms-flux plot is shown in Figure-\ref{fig8}, fitted with a linear function of slope 0.57$\pm$0.1. The Spearman's rank correlation coefficient ($r_{s}$) and the correlation probability (p) are obtained between $<F>$ and $\sigma_{XS}$, which results $r_{s}$ = 0.78 and p = $4.5\times10^{-3}$, suggesting linear rms-flux relation. The obtained lognormal feature in the flux distribution and the linearity between the excess variance to the flux implies the variation in flux is lognormal.

	%FIGURES########################
	%}

%#############################

\section{Summary}
In the present study, we comment on the variability and the spectral characteristics of the blazar TXS 1700+685 using a two-component external Compton model along with SSC for the first time. For the broadband spectral modeling, we focus on the flaring phase for which the simultaneous X-ray observation is present, and the emission region is found to be within BLR. The high energy peak of the broadband SED corresponding to the flaring state is mainly reproduced by the EC process of seed photons originating from the BLR and DT, and the contribution from the disk is found to be sub-dominant. The power deposited by the jet is found to be primarily contributed by the leptonic component, instead of the magnetic field component. \par 

It would be really interesting to undertake the multi-frequency correlation and spectral investigation of this source over several outbursts in order to understand the dominance of leptonic/hadronic processes in different phases of its evolution. It would be interesting to follow up on this source using radio observation to detect any possible micro-lensing effects and connect it with self-correlation studies across its evolution. We strongly encourage the detailed multi-wavelength follow-up of this source in order to uncover such stimulating effects. 

\section{Acknowledgements}
This work made use of data collected by \textit{Fermi} and the \textit{Swift} missions, as well as the archival radio data obtained from \textsc{SEDbuilder} tool. A.B. acknowledges Jayant Abhir for productive discussions regarding the \textit{Fermi} analysis, and Andrea Tramacere for his helpful suggestions regarding multi-wavelength modelling.
DB acknowledges Science and Engineering Research Board - Department of Science and Technology for Ramanujan Fellowship - SB/S2/ RJN-038/2017. For this work, PN acknowledges CSIR fellowship and the support from Physical Research Laboratory which is funded by the Department of Space, India. RP acknowledges the support by the Polish Funding Agency National Science Centre, project 2017/26/A/ST9/00756 (MAESTRO 9), and MNiSW grant DIR/WK/2018/12.

\section{Data Availability}
This work has made use of data from \textit{Fermi}-LAT, \textit{Swift}-XRT, \textit{Swift}-UVOT and archival data from SED builder tool. All the data used are available in the public domain. Details are provided in Section 2. 
%%%%%%%%%%%%%%%%%%%%%%%%%%%%%%%%%%%%%%%%%%%%%%%%%%

\bibliographystyle{mnras}
\bibliography{references} % if your bibtex file is called example.bib
\end{document}